\makeatletter
\disable@package@load{breakurl}{}
\makeatother

\documentclass[sn-mathphys]{sn-jnl}

\jyear{2024}%

\theoremstyle{thmstyleone}%
\newtheorem{theorem}{Theorem}[section]
\newtheorem{corollary}[theorem]{Corollary}%
\newtheorem{lemma}[theorem]{Lemma}%
\newtheorem{problem}[theorem]{Problem}%
\newtheorem{proposition}[theorem]{Proposition}%

\theoremstyle{thmstyletwo}%
\newtheorem{example}[theorem]{Example}%

\theoremstyle{thmstylethree}%
\newtheorem{definition}[theorem]{Definition}%

\usepackage{verbatim}
\newcommand{\bR}{{\mathbb R}}
\newcommand{\bZ}{\mathbb Z}
\newcommand{\al}{\alpha}

\newcommand{\ga}{\gamma}

\newcommand{\si}{\sigma}

\newcommand{\om}{\omega}

\newcommand{\birth}{\mathrm{birth}}
\newcommand{\death}{\mathrm{death}}

\newcommand{\MST}{\mathrm{MST}}

\newcommand{\PD}{\mathrm{PD}}

\newcommand{\bs}{\hfill $\blacksquare$}
\newcommand{\bd}{\partial}

\newcommand{\cech}{\v{C}ech }
\newcommand{\Cech}{\mathrm{\check{C}ech}}

\newcommand{\VR}{\mathrm{VR}}
\newcommand{\Del}{\mathrm{Del}}

\begin{document}

\title[Finite metric spaces with identical or trivial 1D persistence]{Generic families of finite metric spaces with identical or trivial 1-dimensional persistence}


\author[1]{\fnm{Philip} \sur{Smith}}\email{philip.smith3@liverpool.ac.uk}

\author*[1]{\fnm{Vitaliy} \sur{Kurlin}}\email{vitaliy.kurlin@gmail.com}

\affil*[1]{\orgdiv{Department of Computer Science}, \orgname{University of Liverpool}, \orgaddress{\street{Ashton street}, \city{Liverpool}, \postcode{L69 3BX}, \country{United Kingdom}}}


\abstract{
Persistent homology is a popular and useful tool for analysing finite metric spaces, revealing features that can be used to distinguish sets of unlabeled points and as input into machine learning pipelines. 
The famous stability theorem of persistent homology provides an upper bound for the change of persistence in the bottleneck distance under perturbations of points, but without giving a lower bound.
\medskip
 
This paper clarifies the possible limitations persistent homology may have in distinguishing finite metric spaces, which is evident for non-isometric point sets with identical persistence. 
We describe generic families of point sets in metric spaces that have identical or even trivial one-dimensional persistence. 
The results motivate stronger invariants to distinguish finite point sets up to isometry.
}




\maketitle

\section{Motivations, problems, and outline of results}
\label{sec:intro}

Topological Data Analysis (TDA) summarises geometric and topological features in unstructured data and was pioneered by
Serguei Barannikov \cite{barannikov1994framed}, Claudia Landi \cite{frosini1999size}, 
Vanessa Robins \cite{robins1999towards}, 
and Herbert Edelsbrunner et al. \cite{edelsbrunner2000topological}.
The key papers of Gunnar Carlsson \cite{carlsson2009topology}, Robert Ghrist \cite{ghrist2008barcodes}, and Shmuel Weinberger \cite{weinberger2011persistent} were followed by substantial developments of Fred Chazal \cite{chazal2016structure} and others.
\medskip

The main tool of TDA \cite{EdHa08} is persistent homology, which is defined below via a filtration of complexes on a \emph{point cloud} (a finite set $A$ of unordered points).
One can also consider filtrations of sublevel sets of a scalar function.

\begin{definition}[A filtration of complexes $\{C(A;\al)\}$]
\label{dfn:filtration}
Let $A$ be any finite set.
\medskip

\noindent
\textbf{(a)}
A simplicial \emph{complex} $C$ on $A$ is a finite set of subsets $\si\subset A$ (\emph{simplices}) such that all subsets of $\si\subset A$  and hence all intersections of simplices are simplices of $C$.
\medskip

\noindent
\textbf{(b)}
The \emph{dimension} of a simplex $\si$ on $k+1$ points is $k$.
We assume that all points of $A$ are 0-dimensional simplices, sometimes called \emph{vertices} of $C$.
A 1-dimensional simplex (or \emph{edge}) $e$ between points $p,q\in A$ is the unordered pair denoted as $[p,q]$.
\medskip

\noindent
\textbf{(c)}
An ascending \emph{filtration} $\{C(A;\al)\}$ is a family of complexes on the vertex set $A$, paremetrised by a \emph{scale} $\al\in\bR$ so that 
$C(A;\al')\subseteq C(A;\al)$ for $\al'\leq\al$.
\bs
\end{definition}

\begin{definition}[1D persistence and barcode]
\label{dfn:persistence}
For any filtration $\{C(A; \al)\}$ of complexes on a cloud $A$ in a metric space, a homology class $\ga \in H_1(C(A; \al_i))$ is \textit{born} at $\al_i = \birth(\ga)$ if $\ga$ is not in the full image under the induced homomorphism $H_1(C(A; \al)) \to H_1(C(A; \al_i))$ for any $\al < \al_i$. 
The class $\ga$ \textit{dies} at $\al_j = \death(\ga) \geq \al_i$ when the image of $\ga$ under $H_1(C(A; \al_i)) \to H_1(C(A; \al_j))$ merges into the image under $H_1(C(A; \al)) \to H_1(C(A; \al_j))$ for some $\al < \al_i$.
\smallskip

Let $\al_1, \dots, \al_m$ be all scales when a homology class is born or dies in $H_1(C(A; \al))$. 
Let $\mu_{ij}$ be the number of independent classes in $H_1(C(A; \al))$ that are born at $\al_i$ and die at $\al_j$. 
The 1D \textit{persistence} diagram $\PD_1\{C(A; \al)\}\subset\bR^2$ is the multi-set consisting of the pairs $(\al_i, \al_j)$ with integer multiplicities $\mu_{ij}\geq 1$. 
The 1D\emph{barcode} 
is the unordered multi-set of intervals $[\al_i, \al_j)$ with multiplicities $\mu_{ij}$.
\bs
\end{definition}

The birth-death pairs from Definition~\ref{dfn:persistence} can be similarly defined for any $k$-dimensional homology groups $H_k$ with $k\geq 0$ and coefficients in a field, though the coefficients in $\bZ_2=\{0,1\}$ are used in practice and in this 
paper.
\medskip

\begin{figure}[h]
\includegraphics[width=\textwidth]{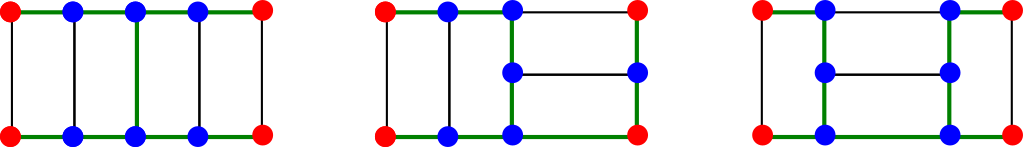}
\caption{
Many non-isometric sets have the same 0D persistence and trivial 1D persistence.
Theorem~\ref{thm:tail} extends these examples to generic families of sets by adding `tails' at red corners. 
}
\label{fig:dominoes}
\end{figure}

Standard filtrations of (Vietoris-Rips, \cech and Delaunay) complexes on a point cloud $A$ are introduced in Definition~\ref{dfn:complexes}.
For all these filtrations in dimension 0, the homology group $H_0(C(A;{\al}))$ is generated by the single-linkage clusters of $A$ formed by all points that can be connected through inter-point distances up to $2\al$.
Then all homology classes of $H_0(C(A;\al))$ are born at $\al=0$ (isolated dots) and die at $\al$ equal to half-lengths of all edges in a Minimum Spanning Tree $\MST(A)$.
Fig.~\ref{fig:dominoes} shows the edges of $\MST(A)$ in green: one edge of length 2 and eight edges of length 1 for each vertex set $A$.

In all examples of Fig.~\ref{fig:dominoes}, the 1-dimensional persistence is trivial (empty due to no pairs with $\death>\birth$) because all rectangular `dominoes' do not create cycles in the 1D homology for the standard filtrations of complexes.
\medskip

In any dimension for filtrations based only on inter-point distances, the resulting persistence diagram is invariant under \emph{isometry} preserving inter-point distances, not up to more general continuous deformations.
\medskip

Hence most persistence-based classifications distinguish point clouds only up to isometry, which is an important equivalence due to the rigidity of many real-life structures.
Fig.~\ref{fig:dominoes} shows sets $A\subset\bR^2$ whose points (in blue and red) form $1\times 2$ `dominoes' that have identical persistence in dimensions 0 and 1.
\medskip

To understand the strength of persistence as an isometry invariant, the following problem asks to fully describe the inverse of the persistence map.

\begin{problem}[Inverting persistence]
\label{pro:inverse}
For a given filtration of complexes, find necessary and sufficient conditions for finite metric spaces to have a given persistence diagram in each dimension.
In particular, describe all \emph{1D homologically trivial} point sets that (by definition) have a trivial (empty) 1D persistence diagram.
\bs  
\end{problem}

\begin{figure}[h!]
\includegraphics[width=\textwidth]{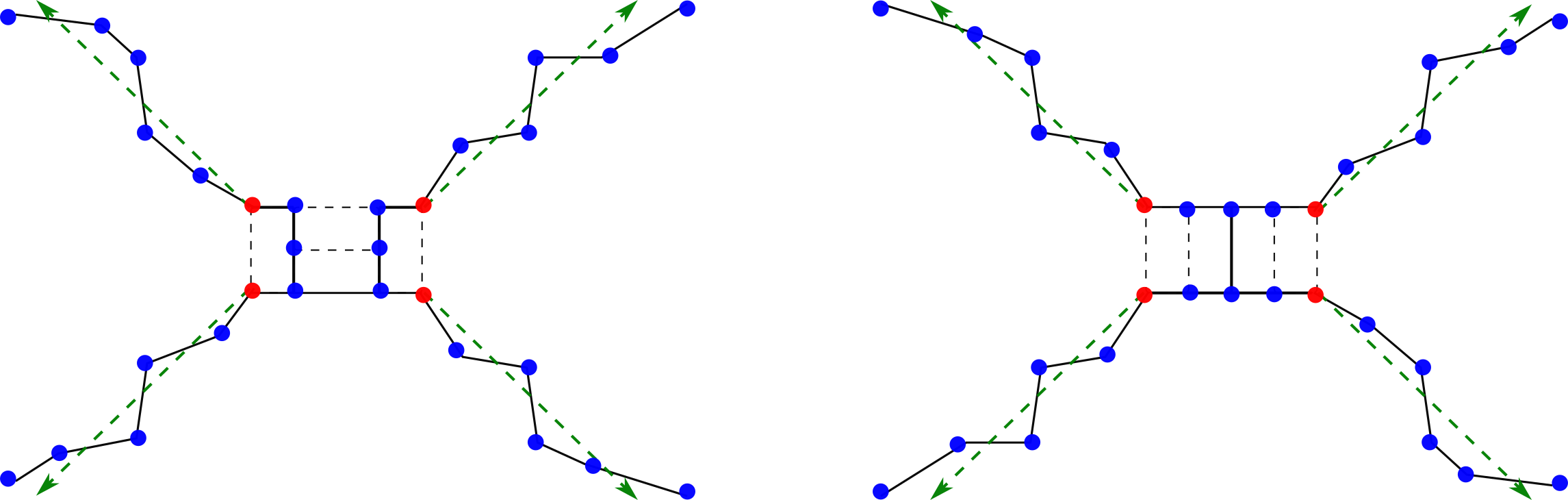}
\caption{
The set $A$ of 10 points in the centre is extended by four tails going out from red points. All such sets have trivial 1D persistence by Corollary~\ref{cor:PD=0}, but all such sets in general position are not isometric to each other.
The black edges form a Minimum Spanning Tree.
}
\label{fig:set+tail}
\end{figure}

The analogue of Problem~\ref{pro:inverse} was solved for 0-dimensional persistence of Morse-like functions on the interval \cite{curry2018fiber}, see also \cite{catanzaro2020moduli}.
Main Theorem~\ref{thm:tail} will show how any number of points can be added to any finite point set whilst leaving the 1-dimensional persistence unchanged, see Fig.~\ref{fig:set+tail} extending 
Fig.~\ref{fig:dominoes}.
\medskip

Corollary~\ref{cor:PD=0} describes generic families of finite metric spaces that have trivial 1-dimensional persistence for the standard filtrations of simplicial (Vietoris-Rips, \cech and Delaunay) complexes in Definition~\ref{dfn:complexes}. 
For high-dimensional data, usually only the Vietoris-Rips complex (determined by its 1D skeleton) is computationally feasible \cite{Bau21}.
Problem~\ref{pro:inverse} for point clouds and persistence in dimensions more than 1 will be discussed in future work.
\medskip

In the context of Problem~\ref{pro:inverse}, the resulting families of point clouds in $\bR^N$ form vast open subspaces (in the space of isometry classes of all point clouds in $\bR^N$), which are mapped by to a single 1D persistence diagram.
This result complements the famous stability theorem \cite{CEH05} stating that under bounded noise, the bottleneck distance between persistence diagrams of a point set and its perturbation has an upper bound depending on the magnitude of the perturbation. 
However, there is no lower bound, so a perturbation of a point set can result in the corresponding persistent homology remaining unchanged. 
\medskip

Section~\ref{sec:persistence} introduces definitions and proves auxiliary lemmas needed for our main Theorem~\ref{thm:tail}, which describes how, given a finite point set, we can add an arbitrarily large point set without affecting the one-dimensional persistent homology.
Section~\ref{sec:experiments} summarises large-scale experiments that reveal important information on the prevalence, or more likely lack, of significant persistent features occurring in randomly generated point clouds in many dimensions.
\medskip

Since persistence is preserved under small perturbations of many point clouds, we might be interested in stronger isometry invariants discussed in section~\ref{sec:discussion}.
Indeed, many applications \cite{edelsbrunner2021density} need to reliably distinguish point sets up to isometry or similar equivalence relations such as rigid motion or uniform scaling. 
A uniform scaling also scales persistence, but a more general continuous deformation of data changes persistence rather arbitrarily.


\section{Edges that are important for 1D persistence}
\label{sec:edges}

This section introduces three classes of edges (short, medium, and long) that will help build point sets with identical 1D persistence.  
Since persistent homology can be defined for any filtration of simplicial complexes on an abstract finite set $A$, the most general settings are recalled in Definition~\ref{dfn:filtration}.
Definition~\ref{dfn:complexes} introduces Vietoris-Rips, \cech and Delaunay complexes on a finite set $A$ in any metric space $M$ or for $A\subset\bR^N$ for Delaunay complexes.
\medskip
  
Let $M$ be any metric space with a distance $d$ satisfying all metric axioms.
An example of a metric space is $\bR^N$ with the Euclidean metric. 
For any points $p,q\in A\subset M$, the edge $e=[p,q]$ has the length $d(p,q)$.
For a point cloud $A\subset\bR^N$, $e=[p,q]$ has the Euclidean length $\vert p-q\vert$ and can be geometrically interpreted as the straight-line segment connecting the points $p,q\in A\subset\bR^N$.
\medskip

Definition~\ref{dfn:complexes} introduces the simplicial complexes 
$\VR(A;\al)$ and $\Cech(A;\al)$ on any finite set $A$ inside an ambient metric space $M$, although $A=M$ is possible.
For a point $p\in A$ and $\al\geq 0$, let $\bar B(p;\al)\subset M$ denote the closed ball with centre $p$ and radius $\al$.
A Delaunay complex $\Del(A;\al)\subset\bR^N$ will be defined for a finite set $A$ only in $\bR^N$ because of extra complications arising if a point set $A$ lives in a more general metric space \cite{boissonnat2018obstruction}.  

\begin{definition}[Complexes $\VR$, $\Cech$, and $\Del$]
\label{dfn:complexes}
Let $A$ be any finite set in a metric space $M$.
Fix a \emph{scale} $\al\geq 0$.
Each complex $C(A;\al)$ below has the vertex set $A$.
\medskip

\noindent
\textbf{(a)}
The \emph{Vietoris-Rips} complex $\VR(A;\al)$ has all simplices on points $p_1,\dots,p_k\in A$ whose pairwise distances are at most $2\al$, so $d(p_i,p_j)\leq 2\al$ for $i\neq j$ in $\{1,\dots,k\}$.
\medskip

\noindent
\textbf{(b)}
The \emph{\cech} complex $\Cech(A;\al)$ has all simplices on points $p_1,\dots,p_k\in A$ such that the full intersection $\cap_{i=1}^k \bar B(p_i;\al)$ is not empty.
\medskip

\noindent
\textbf{(c)}
For any finite set of points $A\subset\bR^N$, the convex hull of $A$ is the intersection of all closed half-spaces of $\bR^N$ containing $A$.
Each point $p_i\in A$ has the \emph{Voronoi domain} 
$$V(p_i)=\{q\in\bR^N\mid \vert q-p_i\vert \leq \vert q-p_j\vert \text{ for any point } p_j\in A, \; p_j\neq p_i\}.$$
The \emph{Delaunay} complex $\Del(A;\al)$ has all simplices on points $p_1,\dots,p_k\in A$ such that $\cap_{i=1}^k (V(p_i)\cap\bar B(p_i;\al))\neq\emptyset$ \cite{delaunay1934sphere}.
Alternatively, a simplex $\si$ on points $p_1,\dots,p_k\in A$ is called a \emph{Delaunay} simplex 
if 
there is an $(N-1)$-dimensional sphere $S^{N-1}$ that passes through the points $p_1,\dots,p_k$ and does not enclose any points of $A$ \cite{shewchuk2000sweep}.
\smallskip

In a degenerate case, the smallest $(k-2)$-dimensional sphere $S^{k-2}$ above can contain more than $k$ points of $A$.
If $\si$ is enlarged to the convex hull $H$ of all points in $A\cap S^{k-2}$, then $\Del(A;\al)$ becomes a polyhedral \emph{Delaunay mosaic}. 
\smallskip

For simplicity, we can choose any triangulation of $H$ into Delaunay simplices.
When the scale $\al$ becomes too large, $\Del(A;\al)\subset\bR^N$ stops growing and becomes a \emph{Delaunay triangulation} of the convex hull of $A$, which is unique in general position.
\medskip

\noindent
The complexes of the types above will be called \emph{geometric} complexes for brevity.
\bs
\end{definition}

Both complexes $\VR(A;\al)$ and $\Cech(A;\al)$ are abstract and so are not embedded in $\bR^N$, even if $A\subset\bR^N$.
Though $\Del(A;\al)$ is embedded into $\bR^N$, its construction has a near-linear time or quadratic time in the size of $A$ only in dimensions $N=2,3$ \cite{cignoni1998dewall}.
For high dimensions $N>3$ or any metric space $M$, the simplest complex to build and store is $\VR(A;\al)$.
Indeed, the Vietoris-Rips complex $\VR(A;\al)$ is a flag complex determined by its 1-dimensional skeleton $\VR^1(A;\al)$ so that any simplex of $\VR(A;\al)$ is built on a complete subgraph whose vertices are pairwise connected by edges in $\VR^1(A;\al)$.  
\medskip

The key idea of persistence is to view any point cloud $A\subset\bR^N$ through 
the lens of a variable scale $\al\geq 0$.
When the scale $\al$ is increasing from the initial value 0, we can form a new topological space from $A$ by replacing points with closed balls of a radius $\al$. 
Then persistent homology identifies topological features of these spaces that 'persist' over a long interval of the scale $\al$.
\medskip

More formally, for any fixed scale $\al\geq 0$, the union $\cup_{p\in A}\bar B(p;\al)$ of closed balls is homotopy equivalent to the \cech complex $\Cech(A;\al)$ and also to the Delaunay complex $\Del(A;\al)\subset\bR^N$ by the Nerve Lemma \cite[Corollary 4G.3]{Hat01}, see also the persistent version in \cite[Lemma 3.4]{chazal2008towards}. 
\medskip

For any geometric complex $C(A;\al)$ from Definition~\ref{dfn:complexes}, all connected components of $C(A;\al)$ are in a 1-1 correspondence with all connected components of the union $\cup_{p\in A}\bar B(p;\al)$ of the closed balls centred at all $p\in A$.
If an edge $e=[p,q]$ enters a complex $C(A;\al)$ at a scale $\al$, then $\al=d(p,q)/2$.
\medskip

Definition~\ref{dfn:edges} makes sense for any filtration of simplicial complexes from Definition~\ref{dfn:filtration}, not only for geometric complexes from Definition~\ref{dfn:complexes}.

\begin{definition}[Short, medium, long edges in a filtration]
\label{dfn:edges}
Let $\{C(A;\al)\}$ be any filtration of complexes on a finite vertex set $A$, see Definition~\ref{dfn:filtration}.
Let an edge $e=[p,q]$ between points $p,q\in A$ enter the complex $C(A;\al)$ at the scale $\al=d(p,q)/2$.
\medskip

\noindent
\textbf{(a)}
Consider the 1-dimensional graph $C'(A;\al)$ with vertex set $A$ and all edges from $C(A;\al)$ except the edge $e$.
If the endpoints of $e$ are in different connected components of $C'(A;\al)$, then the edge $e$ is called \emph{short} in the filtration $\{C(A;\al)\}$.
\medskip

\noindent
\textbf{(b)}
The edge $e$ is called \emph{long} in $\{C(A;\al)\}$ if $A$ has a vertex $v$ such that $C(A;\al)$ has the 2-simplex $\triangle pqv$ and both edges $[p,v],[v,q]$ are in $C(A;\al')$ for some $\al'<\al$.
\medskip

\noindent
\textbf{(c)}
If $e$ is neither short nor long, then the edge $e$ is called \emph{medium} in $\{C(A;\al)\}$.
\bs
\end{definition}

Definition~\ref{dfn:edges}(b) implies that any long edge enters $C(A;\al)$ with a 2-simplex $\triangle pqv$ at the same scale $\al$ and the boundary of this 2-simplex is homologically trivial in $C(A;\al)$ due to the other two edges $[p,v]$ and $[v,q]$ that entered the filtration at a smaller scale $\al'<\al$.

\begin{lemma}[Classes of edges]
\label{lem:classes_edges}
For any finite set $A$ and a filtration $\{C(A;\al)\}$ from Definition~\ref{dfn:complexes}, all edges are split into disjoint classes: short, medium, long.  
\bs
\end{lemma}
\begin{proof}
By Definition~\ref{dfn:edges}(b), the endpoints $p,q$ of any long edge $e=[p,q]\subset C(A;\al)$ are connected by a chain of two edges $[p,v]\cup[v,q]$ that entered the filtration at a smaller scale $\al'<\al$.
Hence the long edge $e$ cannot be short by Definition~\ref{dfn:edges}(a).
So the three classes of edges in Definition~\ref{dfn:edges} are disjoint.
\end{proof}

Definition~\ref{dfn:edges} defined classes of edges for any filtration of complexes.
Proposition~\ref{prop:long_edges} interprets long edges in VR and Cech filtrations via distances.

\begin{proposition}[Long edges in $\VR$, $\Cech$, $\Del$]
\label{prop:long_edges}
Let $A$ be a finite metric space.
\medskip

\noindent
\textbf{(a)}
An edge $e=[p,q]$ in the Vietoris-Rips filtration $\{\VR(A;\al)\}$,  is long if and only if $A$ has a point $v$ such that $e=[p,q]$ is strictly longest in the 2-simplex $\triangle pqv$.
\medskip

\noindent
\textbf{(b)}
An edge $e=[p,q]$ in the \cech filtration $\{\Cech(A;\al)\}$ is long if and only if $A$ has a point $v$ such that $e=[p,q]$ is strictly longest in the 2-simplex $\triangle pqv$ and the triple intersection $\bar B(p;\al)\cap\bar B(q;\al)\cap\bar B(v;\al)$ is not empty for $\al=d(p,q)/2$.
\medskip

\noindent
\textbf{(c)}
For $A\subset\bR^N$,  an edge $e=[p,q]$ in the Delaunay filtration $\{\Del(A;\al)\}$ is long if and only if $A$ has a point $v$ such that $e=[p,q]$ is strictly longest in the 2-simplex $\triangle pqv$ and $V(p)\cap\bar B(p;\al)\cap V(q)\cap\bar B(q;\al)\cap V(v)\cap\bar B(v;\al)\neq\emptyset$ for $\al=\vert p-q \vert /2$.
\medskip

\noindent
\textbf{(d)}
For $A\subset\bR^N$ and any filtration from Definition~\ref{dfn:complexes} and an edge $[p,q]$ in $C(A;\al)$, if $A$ has a point $v$ such that the angle at $v$ in $\triangle pqv$ is not acute, then $[p,q]$ is long. 
\bs
\end{proposition}
\begin{proof}
For all filtrations from Definition~\ref{dfn:complexes}, an edge $e$ enters $C(A;\al)$ at the scale $\al=d(p,q)/2$.
By Definition~\ref{dfn:edges}(b), a long edge enters $C(A;\al)$ together with a 2-simplex $\triangle pqv$ for some $v\in A$, while the other two edges $[p,v],[v,q]$ entered the filtration at a smaller scale $\al'$, the edge $e=[p,q]$ is longest in the 2-simplex $\triangle pqv$.
\medskip

\noindent
\textbf{(b)}
For the \cech filtration, the triple intersection $\bar B(p;\al)\cap\bar B(q;\al)\cap\bar B(v;\al)$ is non-empty to guarantee that  $\Cech(A;\al)$ includes $\triangle pqv$ by Definition~\ref{dfn:complexes}(b).
\medskip

\noindent
\textbf{(c)}
For the Delaunay filtration, $V(p)\cap\bar B(p;\al)\cap V(q)\cap\bar B(q;\al)\cap V(v)\cap\bar B(v;\al)$ is not empty to guarantee that $\Del(A;\al)$ includes $\triangle pqv$ by Definition~\ref{dfn:complexes}(c).
\medskip

\noindent
\textbf{(d)}
For all filtrations (Vietoris-Rips, \cech and Delaunay) and $A\subset\bR^N$, if the angle at $v$ in $\triangle pqv$ is not acute, then $[p,q]$ is strictly longest in $\triangle pqv$, which finishes the proof for the Vietoris-Rips filtration by part (a).
The closed ball $\bar B(u;\al)$ centred at the mid-point $u$ of $[p,q]$ contains all $p,q,v$, so the point $u$ belongs to $\bar B(p;\al)\cap\bar B(q;\al)\cap\bar B(v;\al)$, which finishes the proof for the \cech filtration by part (b).
\smallskip

For the Delaunay filtration, since the edge $[p,q]$ entered $\Del(A;\al)$ at the scale $\al$, Definition~\ref{dfn:complexes}(c) gives an $(N-1)$-dimensional sphere $S^{N-1}$ that passes through $p,q$ and does not enclose any point of $A$.
Let $S(v)$ be the smallest $(N-1)$-dimensional sphere that passes through $p,q,v$.
If $S(v)$ encloses (strictly inside) no points of $A$, then the 2-simplex $\triangle pqv$ is Delaunay by Definition~\ref{dfn:complexes}(c) and enters $\Del(A;\al)$ together with $[p,q]$ at $\al=\vert p-q \vert /2$, so $[p,q]$ is long by Definition~\ref{dfn:edges}(b).
Otherwise, we will find another empty sphere circumscribing a non-acute Delaunay 2-simplex on $[p,q]$.  
\smallskip

The centres of the spheres $S^{N-1}$ and $S(v)$ lie in the $(N-1)$-dimensional hyperspace $H$ that perpendicularly splits the edge $[p,q]$ at its mid-point $u$.
Connect these centres by the straight-line path of points $O_t$, $t\in[0,1]$, within $H$. 
For every centre $O_t\in H$, consider the $(N-1)$-dimensional sphere $S_t$ with the radius $R_t=\vert O_t -p \vert =\vert O_t -q \vert$ so that $S_t$ passes through $p,q$ for all $t\in[0,1]$, see Fig.~\ref{fig:edges}~(left).

\begin{figure}[h!]
\includegraphics[height=20mm]{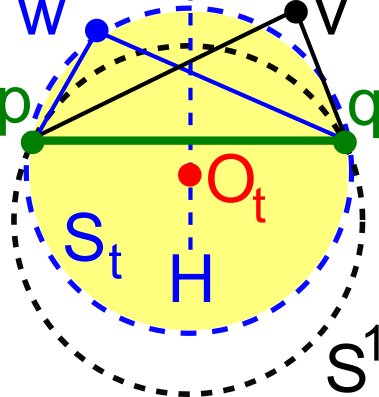}
\hspace*{5mm}
\includegraphics[height=20mm]{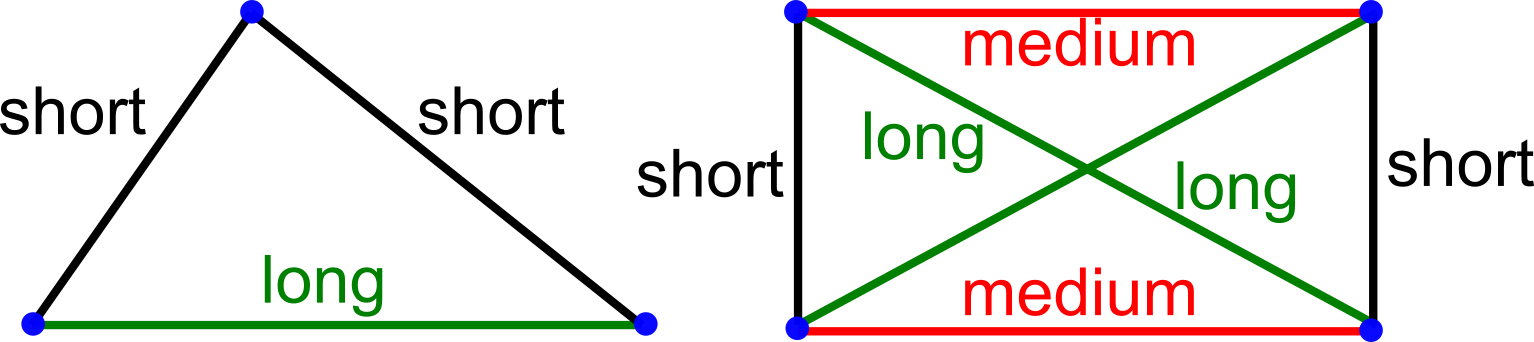}
\caption{
\textbf{Left}: an edge $[p,q]$ opposite to a non-acute angle in a 2-simplex $\triangle pqw$, see the proof of Proposition~\ref{prop:long_edges}(d).
\textbf{Middle} and \textbf{Right}: classes of edges by Definition~\ref{dfn:edges} in Example~\ref{exa:edges}.}
\label{fig:edges}
\end{figure}

Then the continuous family of spheres deforming from $S^{N-1}$ to $S(v)$ should contain a sphere $S_t$ that passes through a point $w\in A-\{p,q\}$ and encloses no points of $A$.
This point $w$ should lie inside the spherical segment bounded by $S(v)$ and the $(N-1)$-dimensional hyperspace $H_1$ passing through $[p,q]$ orthogonally to $[u,O_1]$.
\smallskip

Since this segment is not larger than a half-ball bounded by $S(v)$, any such point $w$ has a non-acute angle $\angle pwq$ on the diameter $[p,q]$ of the $(N-2)$-dimensional sphere $S(v)\cap H_1$.
Then the non-acute 2-simplex $\triangle pqw$ is Delaunay by Definition~\ref{dfn:complexes}(c) and enters $\Del(A;\al)$ together with $[p,q]$, so $[p,q]$ is long by Definition~\ref{dfn:edges}(b).
\end{proof}

\begin{example}[Classes of edges on 3 and 4 points]
\label{exa:edges}
For any 3-point set $A\subset\bR^N$, let the edges of $A$ have lengths $\vert e_1 \vert \leq \vert e_2 \vert < \vert e_3 \vert$. 
By Definition~\ref{dfn:edges}, in $\{\VR(A;\al)\}$ the edge $e_3$ is long whilst the edges $e_1,e_2$ are short, see Fig.~\ref{fig:edges}~(middle).
If $\vert e_1 \vert < \vert e_2 \vert = \vert e_3 \vert$, then the edge $e_1$ is short but both edges $e_2,e_3$ are medium, not long.
If $\vert e_1 \vert = \vert e_2 \vert = \vert e_3 \vert$, then all three edges are medium.
Let $C(A;\al)$ be any geometric complex from Definition~\ref{dfn:complexes} on a finite set $A\subset\bR^2$.
If the set $A$ consists of four vertices of the unit square, all square sides are medium whilst both diagonals are long, see Fig.~\ref{fig:edges}~(right). 
If the set $A$ consists of four vertices of a rectangle that is not a square, the two shorter sides are short, the longer sides are medium and both diagonals are long. 
\bs
\end{example}

\section{Tails without medium edges in a metric space}
\label{sec:tails}

As usual, we consider homology groups with coefficients in a field, say $\bZ_2$. 
  
\begin{proposition}[No medium edges $\Rightarrow$ trivial $H_1$]
\label{prop:no-medium}
For any filtration $\{C(A;\al)\}$ on a finite set $A$ from Definition~\ref{dfn:filtration}, when a scale $\al\geq 0$ is increasing, a new homology cycle in $H_1(C(A;\al))$ can be created only due to a medium edge in $C(A;\al)$.
Hence, if $\{C(A;\al)\}$ has no medium edges, then $H_1(C(A;\al))$ is trivial for $\al\geq 0$.
\bs
\end{proposition}
\begin{proof}
When building the complex $C(A;\al)$, if we add a short edge $e$, by Definition~\ref{dfn:edges}(a), the previously disjoint components of $C^1(A;\al)$ containing the endpoints $p,q$ of $e$ become connected.
Hence no 1-dimensional cycle in 
$C^1(A;\al)$ is created.
\medskip

 
For any $\al$, let a cycle $\ga$ have just appeared in $H_1(C(A;\al))$, represented by several edges including $e_1,...e_k$ that have appeared at the same scale $\al$. 
By Lemma~\ref{lem:classes_edges} each $e_i$ is either short, medium or long. 
By Definition~\ref{dfn:edges}(b) any long edge $e=[p,q]$ enters $C(A;\al)$ strictly after two shorter edges $[p,v],[v,q]$, and at the same time as the triangle $\triangle pqv$.
The cycle $\ga$ including the edge $[p,q]$ is homologically equivalent to the cycle with $[p,q]$ replaced with the chain $[p,v]\cup[v,q]$.
Hence we can assume that all $e_1,\dots,e_k$ are either short or medium. 
Since the endpoints of $e_i$ are connected by the complementary path $\ga-e_i$, each $e_i$ cannot be short by Definition~\ref{dfn:edges}(a) for $i=1,\dots,k$.
So $\ga$ contains at least one medium edge.
Since only medium edges lead to non-trivial cycles, if $A$ has no medium edges, then $H_1(C(A;\al))$ is trivial.
\end{proof}

\begin{definition}[Tail of points]
\label{dfn:tail}
For a fixed filtration $\{C(A;\al)\}$ on a finite set $A$ from Definition~\ref{dfn:filtration}, a \emph{tail} $T$ in a metric space $M$ is any ordered sequence $T=\{p_1,\dots,p_n\}$, where $p_1$ is the \emph{vertex} of $T$, any edge $[p_i,p_{i+1}]$ between successive points is short, and any edge $[p_i,p_{j}]$ between non-successive points is long for any $1\leq i<j\leq n$.
\bs
\end{definition}

\begin{proposition}[Tails have trivial $\PD_1$]
\label{prop:tail}
Any tail $T$ from Definition~\ref{dfn:tail} for a filtration $\{C(T;\al)\}$ of complexes from Definition~\ref{dfn:filtration} has 
trivial 1D persistence.
\end{proposition}
\begin{proof}
Since any tail $T$ has no medium edges by Lemma~\ref{lem:classes_edges}, the tail $T$ has trivial $H_1(C(T;\al))$ for any $\al\geq 0$ by Proposition~\ref{prop:no-medium}, hence trivial 1D persistence.
\end{proof}

\begin{figure}[h]
\includegraphics[height=17mm]{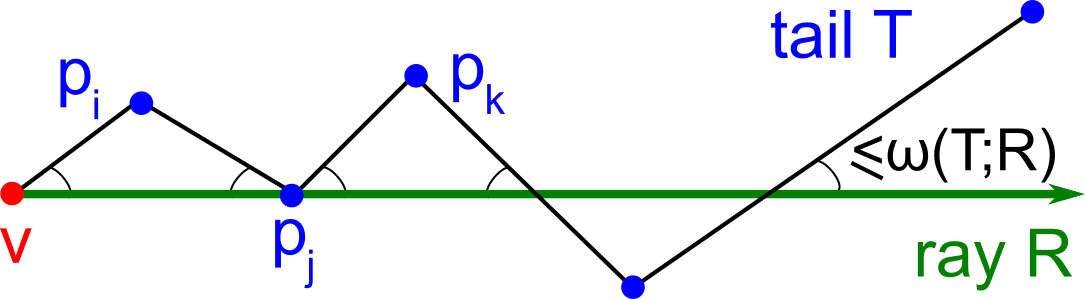}
\hspace*{2mm}
\includegraphics[height=17mm]{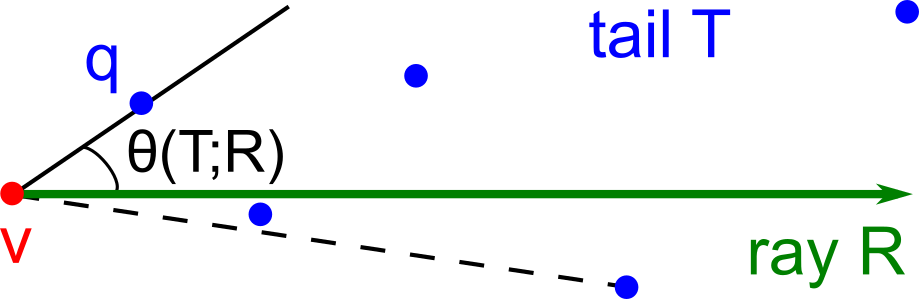}
\caption{
A tail $T$ around a ray $R$ with vertex $v$ in $\bR^2$, see Definitions~\ref{dfn:ang_deviation} and~\ref{dfn:thickness}.
\textbf{Left}: all angles are not greater than the angular deviation $\om(T;R)$.
\textbf{Right}: the angular thickness $\theta(T;R)$ can be smaller than the angular deviation $\om(T;R)$.}
\label{fig:ray+tail}
\end{figure}

If vectors are not explicitly specified, all edges and straight lines are unoriented.
We measure the angle between unoriented straight lines as their minimum angle within $[0,\frac{\pi}{2}]$, see Fig.~\ref{fig:ray+tail}~(left).

\begin{definition}[Angular deviation $\omega(T;R)$ from a ray $R$]
\label{dfn:ang_deviation}
In $\bR^N$, a \emph{ray} is any half-infinite line $R$ going from a point  $v$ (the \emph{vertex} of $R$).
For any sequence $T=\{p_1,\dots,p_n\}$ of ordered points in $\bR^N$, the \emph{angular deviation} $\om(T;R)$ of $T$ relative to $R$ is the maximum angle $\angle(R,[p, q])\in[0,\frac{\pi}{2}]$ over all distinct points $p,q\in T$.
\bs
\end{definition}

\begin{lemma}[Tails in $\bR^N$]
\label{lem:tail}
In $\bR^N$, let $R$ be a straight infinite ray with a vertex $v=p_1$ and $T$ be any sequence of points $p_1,\dots,p_n$ with an angular deviation $\om(T;R)<\frac{\pi}{4}$.
\medskip

\noindent
\textbf{(a)}
For any $i<j<k$, the angle $\angle p_i p_j p_k$ is non-acute.
The edge between the non-successive points $p_i,p_k$ is long in any filtration $\{C(T;\al)\}$ in Definition~\ref{dfn:complexes}.
\medskip

\noindent
\textbf{(b)} 
Any edge between successive points $p_{j-1},p_j$, $j=2,\dots,n$, is short in $\{C(T;\al)\}$.
\medskip

\noindent 
Hence $T$ has no medium edges in $\{C(T;\al)\}$ and is a tail by Definition~\ref{dfn:tail}.
\end{lemma}
\begin{proof}
\textbf{(a)}
The condition $\om(T;R)<\frac{\pi}{4}$ implies that all points of $T$ can be ordered by their distance from the vertex $v=p_1$ to their orthogonal projections in the ray $R$.
Apply a parallel shift to $p_i,p_j,p_k$ so that $p_j\in R$.
In the 2-simplex $\triangle p_i p_j p_k$, the angle 
$$\angle p_i p_j p_k = \pi-\angle(R,[p_j p_i])-\angle(R,[p_j p_k])\geq\pi-2\om(T;R)>\frac{\pi}{2}$$ is non-acute due to $\om(T;R)<\frac{\pi}{4}$, hence strictly largest.
By Proposition~\ref{prop:long_edges}(d) the edge $[p_i,p_k]$ is long in any filtration $\{C(T;\al)\}$ in the sense of Definition~\ref{dfn:edges}(b).
In particular, the edge $[p_i,p_k]$ is longer than both $[p_i,p_j]$ and $[p_j,p_k]$ for any $i<j<k$.
\medskip

\noindent
\textbf{(b)} 
The points $p_{j-1},p_{j}$ remain in disjoint components of $C^1(T;\al)$ after adding all other edges of the same length $\vert p_{j}-p_{j-1} \vert$.
Indeed, we proved above that any other edge connecting non-successive points $p_i,p_k$ for $i\leq j-1<j\leq k$ is longer than the edge $[p_{j-1},p_j]$ between intermediate successive points. 
\end{proof}

Fig.~\ref{fig:ray+tail}~(right) illustrates the angular thickness below for Theorem~\ref{thm:tail} later.
 
\begin{definition}[Angular thickness $\theta$]
\label{dfn:thickness}
Let $R \subset \bR^N$ be a ray with a vertex $v=p_1$, $T=\{p_1,\dots,p_n\}$ be a finite sequence of points.
The \emph{angular thickness} $\theta(T;R)$ of $T$ with respect to $R$ is the maximum angle $\angle(R,[p_1,p_i])$ for $i=2,\dots,n$.
\bs
\end{definition}

\section{Persistence for long wedges and with tails}
\label{sec:persistence}

This section proves main Theorem~\ref{thm:tail} saying that the 1D persistence for a point cloud $A$ remains unchanged under adding a suitable tail $T$ of points to $A$.
The key step is Theorem~\ref{thm:long_wedges} describing how to compute the 1D persistence for a union of point clouds $\cup_{i=1}^k A_i$ sharing a single point as defined below.    

\begin{definition}[A long wedge]
\label{dfn:long_wedge}
Let $A_1,\dots,A_k$ be finite point clouds sharing one common point $v$.
In a filtration $\{C(\cup_{i=1}^k A_i;\al)\}$ from Definition~\ref{dfn:filtration}, call a simplex \emph{heterogeneous} if its vertices don't include $v$ and belong to at least two different clouds $A_i$ for $i=1,\dots,k$.
Assume that any heterogeneous edge of $\{C(\cup_{i=1}^k A_i;\al)\}$ is long in the sense of Definition~\ref{dfn:edges}(b).
Also assume that if any heterogeneous 2-simplex $abc$ enters the filtration $\{C(\cup_{i=1}^k A_i;\al)\}$ at a scale $\al$, then $C(\cup_{i=1}^k A_i;\al)$ includes the 2-simplices $abv$, $bcv$, $cav$.
Then the union $\cup_{i=1}^k A_i$ is called a \emph{long wedge}.
\bs 
\end{definition}


In topology, a \emph{wedge} (or bouqet) $\vee_{i=1}^k C(A_i;\al)$  of complexes, each with a base point $v_j$, is the quotient of the disjoint $\sqcup_{i=1}^k C(A_i;\al)$, where all base points $v_1,\dots,v_k$ are collapsed to one point $v$.
\cite[Corollary 2.25]{Hat01} proves an isomorphism $H_1(\vee_{i=1}^k C(A_i;\al))\to \oplus_{i=1}^k H_1(C(A_i;\al))$.
Theorem~\ref{thm:long_wedges} proves a similar isomorphism for the larger complex  $C(\cup_{i=1}^k A_i;\al)$ of a long wedge $\cup_{i=1}^k A_i$ of point clouds instead of the wedge $\vee_{i=1}^k C(A_i;\al)$ of smaller complexes.

\begin{theorem}[Persistence of a long wedge]
\label{thm:long_wedges}
For any filtration $\{C(\cup_{i=1}^k A_i;\al)\}$ of a long wedge from Definition~\ref{dfn:long_wedge}, 
$H_1(C(\cup_{i=1}^k A_i;\al))$ is isomorphic to the direct sum $\oplus_{i=1}^k H_1(C(A_i;\al))$ for all $\al$.
Hence the 1D persistence diagram $\PD_1\{C(\cup_{i=1}^k A_i;\al)\}$ is the union of the 1D persistence diagrams $\PD_1\{C(A_i;\al)\}$ for $i=1,\dots,k$.
\bs
\end{theorem}
\begin{proof}
Due to the isomorphism $H_1(\vee_{i=1}^k C(A_i;\al))\cong \oplus_{i=1}^k H_1(C(A_i;\al))$ by \cite[Corollary 2.25]{Hat01}, it suffices to prove that
$H_1(\vee_{i=1}^k C(A_i;\al))\cong H_1(C(\cup_{i=1}^k A_i;\al))$.
\smallskip

The inclusion $\vee_{i=1}^k C(A_i;\al)\subset C(\cup_{i=1}^k A_i;\al)$ induces the homomorphism $H_1(\vee_{i=1}^k C(A_i;\al))\to H_1(C(\cup_{i=1}^k A_i;\al))$ whose bijectivity is proved below.
\medskip

\emph{Surjectivity} of $h$. 
By Definition~\ref{dfn:edges}(b) any long edge $e=[p,q]$ belongs to a complex $C(\cup_{i=1}^k A_i;\al)$ together with a 2-simplex $pvq$ whose edges $[p,v]$ and $[q,v]$ have already entered $C(\cup_{i=1}^k A_i;\al')$ for some $\al'<\al$.
\smallskip

Replace the edge $[p,q]$ with the homologous chain $[p,v]\cup[v,q]$ in  $C(\cup_{i=1}^k A_i;\al)$.
Continue applying these replacements for other long edges until any cycle of edges in $C(\cup_{i=1}^k A_i;\al)$ becomes homologous to a sum of non-long edges.
\smallskip

By Definition~\ref{dfn:long_wedge}, both endpoints of any remaining non-long edge in $C(\cup_{i=1}^k A_i;\al)$ belong to the same cloud $A_i$.
Then the resulting cycle is a sum of $k$ sums $s_1,\dots,s_k$, where each $s_i$ is a sum of only edges from $C(A_i;\al)$. 
Since all clouds $A_i$ share a single point, the resulting cycle is a wedge (1-point union) of the sums $s_1,\dots,s_k$, which should be cycles in $C(A_i;\al)$ for $i=1,\dots,k$, respectively.
So any cycle in $H_1(C(\cup_{i=1}^k A_i;\al))$ is homologous to an element in 
$H_1(\vee_{i=1}^k C(A_i;\al))$.
\medskip

\emph{Injectivity} of $h$. 
It remains to prove that if any 1-dimensional cycle $\ga$ in $\vee_{i=1}^k C(A_i;\al)$ is bounded by a 2-dimensional chain $\si\in C(\cup_{i=1}^k A_i;\al)$, then $\ga$ is bounded by a chain $\tau$ in $\vee_{i=1}^k C(A_i;\al)$. 
By Definition~\ref{dfn:long_wedge} replace any heterogeneous 2-simplex $[abc]$ in the closure of $C(\cup_{i=1}^k A_i;\al)-(\vee_{i=1}^k C(A_i;\al))$ with the sum of non-heterogeneous simplices $[abv]+[bcv]+[cav]$, whose total boundary is $\bd[abc]$.
\smallskip

After all such replacements, we get a chain $\tau$ that has the same boundary $\bd\tau=\ga$ and has no heterogeneous simplices.
The boundary $\bd\tau$ also has no heterogeneous edges $[p,q]$ with $p\in A_i-\{v\}$ and $q\in A_j-\{v\}$ for $i\neq j$, else $\ga=\bd\tau$ is not within the wedge $\vee_{i=1}^k C(A_i;\al)$ of complexes.
Hence every 2-simplex of $\tau$ is within a single cloud $A_i$ for some $i=1,\dots,k$, so the whole chain $\tau$ is within $\vee_{i=1}^k C(A_i;\al)$.
\end{proof}

Definition~\ref{dfn:general_pos} is needed by \cite[Theorem 5.10]{bauer2017morse} to guarantee that the filtration of \cech and Delaunay complexes have the same persistence.

\begin{definition}[A cloud in general position]
\label{dfn:general_pos}
A finite cloud $A\subset\bR^N$ is in \emph{general position} if every subset $P\subset A$ of at most $N + 1$ points is affinely independent, and no point of $A-P$ lies on the smallest $(N-1)$-dimensional circumsphere of $P$.
\bs
\end{definition}

Theorem~\ref{thm:tail} can be considered a Euclidean example of Theorem~\ref{thm:long_wedges} and describes sufficient conditions for a cloud $A$ and a tail $T$ to guarantee that three types of filtrations on $A\cup T$ and $A$ have the same persistence $\PD_1$.
\smallskip


\begin{theorem}[A long wedge with a tail]
\label{thm:tail}
Let $A\subset\bR^N$ be a finite set, $v\in A$ be on the boundary of the convex hull of $A$, and $R$ be a ray with a vertex $v$ so that $\mu(R;A)=\min\limits_{p\in A-\{v\}}\angle(R,[v,p])\geq\frac{\pi}{2}$.
Let $T$ be a tail with the vertex $v$ such that $\mu(R;A)\geq\theta(T;R)+\frac{\pi}{2}$ and $A\cup T$ is in general position by Definition~\ref{dfn:general_pos}.
For 
any filtration from Definition~\ref{dfn:complexes},
we have that $\PD_1\{C(A\cup T;\al)\}=\PD_1\{C(A;\al)\}$.
\bs
\end{theorem}
\begin{proof}
Any heterogeneous edge $[p,q]$ with $p\in A$ and $q\in T$ has a non-acute angle at $v$
$$\angle pvq\geq \angle(R,[v,p])-\angle(R,[v,q])
\geq \mu-\angle(R,[v,q])\geq\mu-\theta(T;R)>\frac{\pi}{2}.$$
Due to the point $v$, the heterogeneous edge $[p,q]$ is long in $\{C(A\cup T;\al)\}$  by Proposition~\ref{prop:long_edges}(d).
To prove that $A\cup T$ is a long wedge by Definition~\ref{dfn:long_wedge},
 consider any heterogeneous 2-simplex $abc$ in the complex $\{C(A\cup T;\al)\}$.
In the boundary $\bd[abc]$, any heterogeneous edge, say $[a,b]$, is strictly the longest by the argument above, while the edges $[a,v]$, $[b,v]$ are no longer heterogeneous, so $A\cup T$ is a long wedge.
\smallskip

In the case of a \cech filtration, let $\triangle abc$ be any heterogeneous 2-simplex in $\Cech(A\cup T;\al)$ such that (say) $a\in A$ and $b,c\in T$.
For the heterogenous edges $[a,b]$ and $[a,c]$, the earlier proved inequalities $\angle avb\geq\frac{\pi}{2}$ and $\angle avc\geq\frac{\pi}{2}$ implies that $v$ belongs to the smallest closed circumballs of $[a,b]$ and $[a,c]$, hence to the smallest closed circumball of $\triangle abc$.
Then the 3-simplex $abcv$ and all its faces belong to $\Cech(A\cup T;\al)$.
All conditions of Definition~\ref{dfn:long_wedge} hold, so $A\cup T$ is a long wedge.
\smallskip

Since the tail $T$ has the trivial (empty) 1D persistence by Proposition~\ref{prop:tail}, Theorem~\ref{thm:long_wedges} implies that $\PD_1\{C(A\cup T;\al)\}=\PD_1\{C(A;\al)\}$ for any filtration form Definition~\ref{dfn:complexes}.
By \cite[Theorem 5.10]{bauer2017morse}, the Delaunay and \cech filtrations have the same persistence for clouds in general position, which finishes the proof.
\end{proof}

\begin{figure}[h]
\includegraphics[height=36mm]{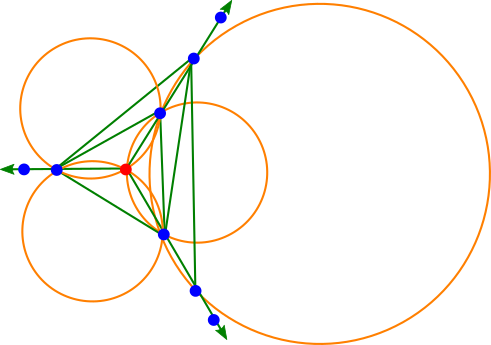}
\hspace*{1mm}
\includegraphics[height=36mm]{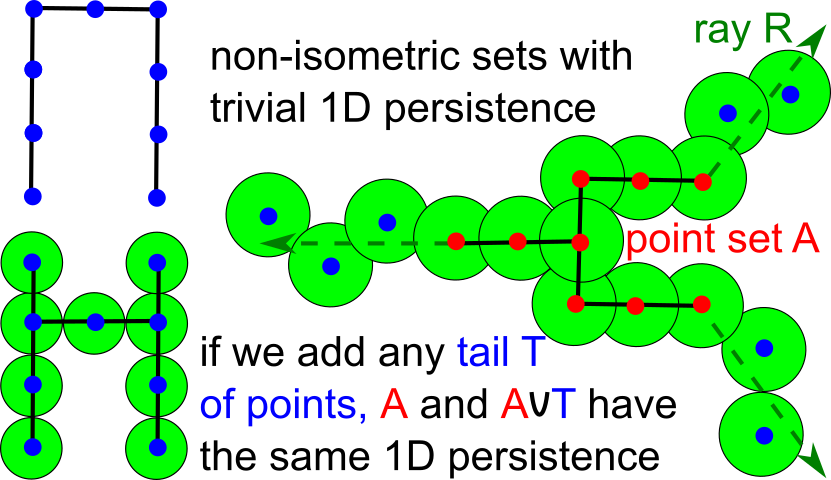}
\caption{
\textbf{Left}: the cloud $A$ in Theorem~\ref{thm:tail} can be a single red point extendable by tails of blue points along straight rays that form non-acute angles. 
Then all Delaunay triangles are obtuse, circumscribed by orange circles, meaning that $\PD_1\{\Del(C;\al)\}=\emptyset$. 
\textbf{Right}: a tail $T$ can be generically perturbed under conditions of Theorem~\ref{thm:tail} without changing $\PD_1$.
}
\label{fig:cloud+tails}
\end{figure}

Fig.~\ref{fig:cloud+tails}~(left) illustrates a Delaunay filtration on a cloud $A\subset\bR^2$.
All blue points lie on rays that have pairwise angles $120^\circ$ and emanate from a red point $v$ so that all green Delaunay triangles are obtuse with all orange circumcircles not enclosing any points of $A$, which implies that $\PD_1\{C(A;\al)\}$ is empty.

\begin{corollary}[Clouds with $\PD_1=\emptyset$]
\label{cor:PD=0}
If a point cloud $A$ has 
$\PD_1\{C(A;\al)\}=\emptyset$, then any long wedge $A\cup T$ with a tail $T$ has $\PD_1\{C(A\cup T;\al)\}=\emptyset$. 
\bs
\end{corollary}
\begin{proof}
Since the tail $T$ has trivial 1D persistence by Proposition~\ref{prop:tail}, Theorem~\ref{thm:long_wedges} implies that 
$\PD_1\{C(A\cup T;\al)\}=\PD_1\{C(A;\al)\}=\emptyset$, see Fig.~\ref{fig:cloud+tails}~(right).
\end{proof}

\section{Experiments on persistence of random sets}
\label{sec:experiments}

The experiments in this section use the Vietoris-Rips filtration whose 1-dimensional persistence is computed by Ripser\cite{Bau21}, a fast implementation of Vietoris-Rips persistence.
The code of the first author is available in \cite{smith2022trivial}.
\medskip

The aim is to understand how often random point sets have trivial persistence or cycles with only low persistence, see more general conjectures  \cite{bobrowski2022universality}.
The experiments depend on two parameters, the size $n$ of a set, and the dimension $N$ that the point set lies in. 
For each $n, N$ in the ranges chosen, we generate 1000 point sets of $n$ points uniformly sampled in a unit $N$-dimensional cube. 
\medskip

\begin{figure}[h]

\centering

\includegraphics[width=\textwidth]{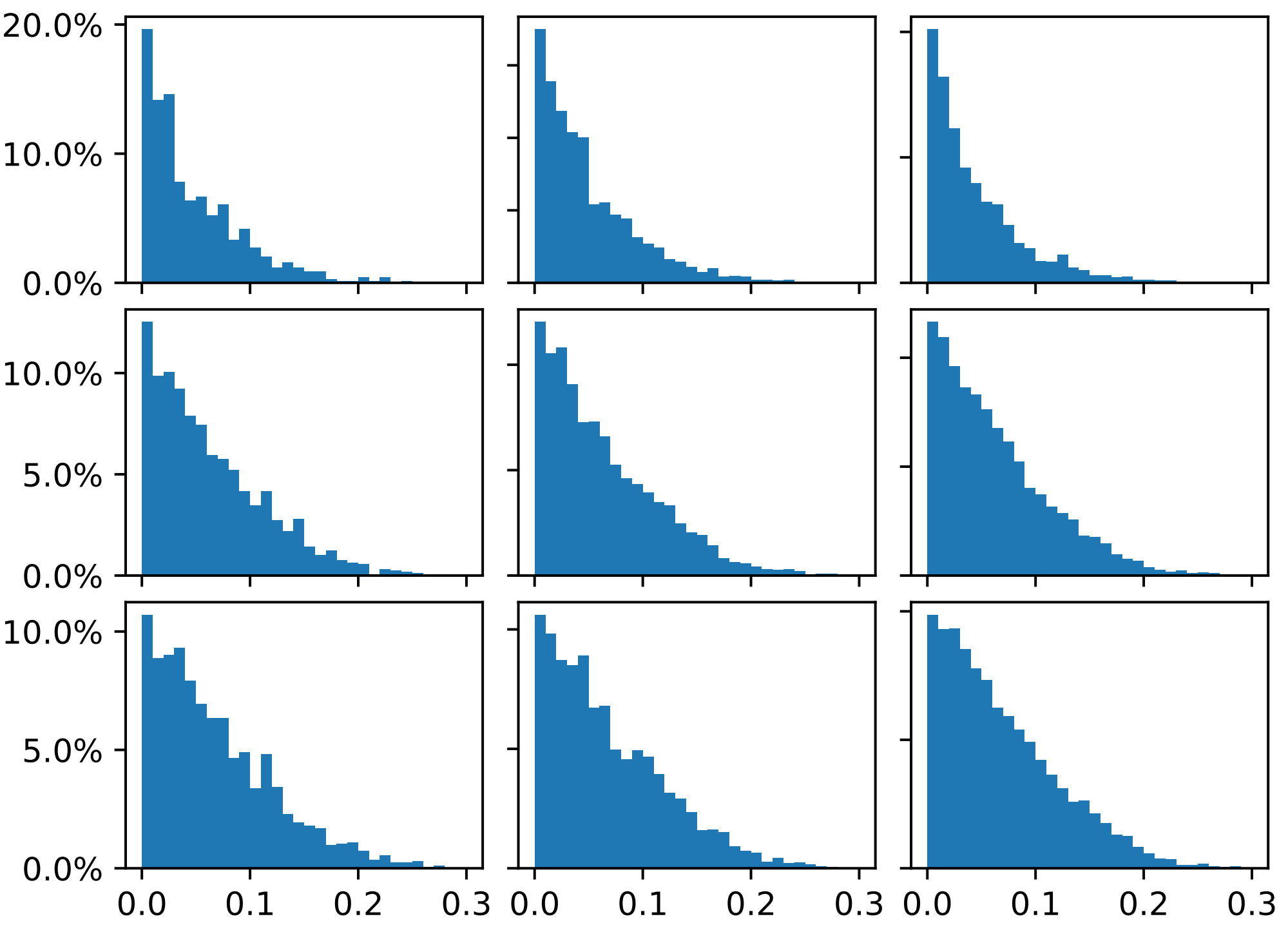}

\caption{Histograms of the persistence $p=$ death$-$birth in 1000 point sets in nine configurations of the parameters $n$ and $N$. 
The $x$-axis is the persistence $p$, the $y$-axis is the percentage of pairs (birth,death) with the given persistence $p$. Top row: $N = 2$; middle row: $N = 5$; bottom row: $N = 8$. Left column: $n = 10$; middle column: $n = 15$; right column $n = 20$. }
\label{fig:hist}
\end{figure}

Figure~\ref{fig:hist} shows histograms of the 1-dimensional persistence (death$-$birth) for nine configurations of the parameters: set sizes $n=10,15,20$ and dimensions $N=2,5,8$.
Each histogram highlights that one-dimensional persistent features are skewed towards a low persistence.
Geometrically, the pairs (birth,death) would be close to the diagonal in a persistence diagram.
\medskip

\begin{figure}[h]
\centering
\includegraphics[width=\textwidth]{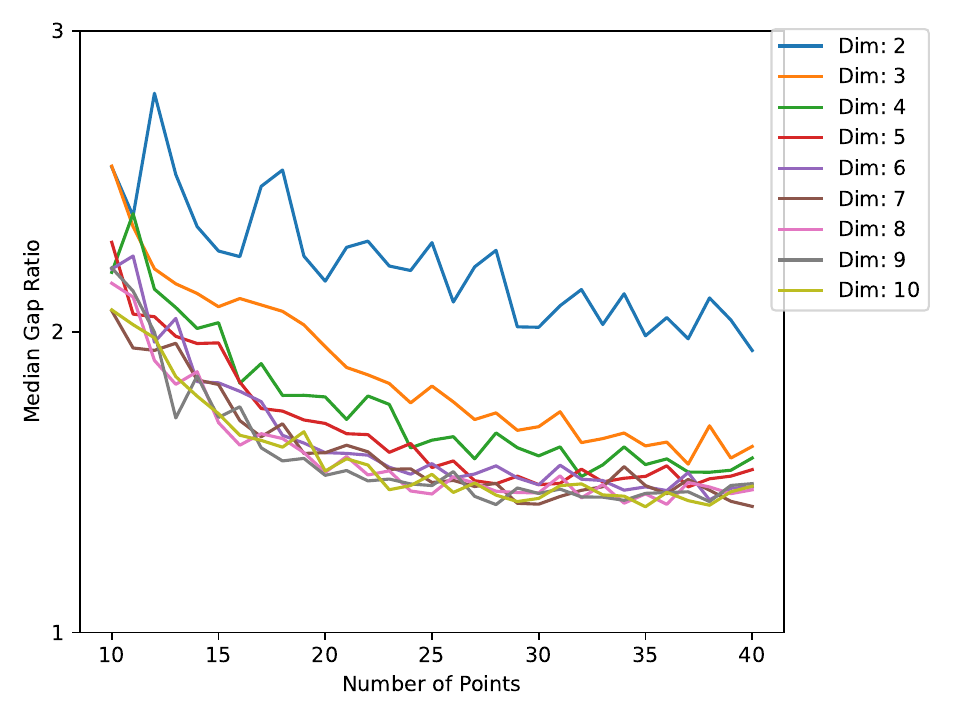}
\caption{The median gap ratio of a point set with at least two 1D persistent features, as the set size varies from $n = 10$ to $n = 40$ and the dimension $N$ varies from $N = 2$ to $N = 10$.}
\label{fig:gr1040}
\end{figure}

Recall that highly persistent features (birth,death) are naturally separated from others with lower persistence $p=$ death$-$birth by the widest diagonal gap in the persistence diagram, see \cite{smith2021skeletonisation}.
If we order all pairs (birth,death) by their persistence $0<p_1\leq\dots\leq p_k$, the widest gap has the largest difference $p_{i+1}-p_i$ over $i=1,\dots,k-1$. 
This widest gap can separate several pairs (birth,death) from the rest, not necessarily just a single feature. 
\medskip

However, the first widest gap is significant only if it can be easily distinguished from the second widest gap.
So the significance of persistence can be measured as the ratio of the first widest gap over the second widest gap.
This invariant up to uniform scaling of given data is called the \emph{gap ratio}.
Figure~\ref{fig:gr1040} shows the median gap ratio calculated over 1000 random point clouds in a unit cube for dimensions $N=2,\dots,10$ and point set sizes $n=10,\dots,40$.
\medskip

Figure~\ref{fig:gr1040} implies that for higher dimensions N, the median gap ratio quickly decreases to within the range [1,2] as the number $n$ of points is increasing.
Hence, for pure random clouds when a persistence diagram contains at least two pairs (birth,death) above the diagonal, it is harder to separate highly persistent features from noisy artefacts that are close to the diagonal.
\medskip

Figure~\ref{fig:gr1040} also seems to suggest a limiting distribution 
as $N\to+\infty$.

\section{Conclusions and discussion of other invariants}
\label{sec:discussion}

Main Theorem~\ref{thm:tail} showed how one can add an arbitrarily long tails to an existing point set without affecting the 1-dimensional persistent homology.
Corollary~\ref{cor:PD=0} implies that families of sets with trivial 1D persistence form vast continuous subspaces in the space of isometry classes of finite sets.
\smallskip

The bottleneck distance between persistence diagrams vanishes on these subspaces and cannot have a lower bound.
We conjecture that Theorem~\ref{thm:tail} extends to any higher-dimensional persistence in the following open problem.

\begin{problem}[Adding tails preserves any persistence]
\label{pro:tail_higher_dim}
Check if, for any point cloud $A\subset\bR^N$ and a tail satisfying Theorem~\ref{thm:tail}, adding the tail $T$ to $A$ preserves any $k$-dimensional persistence, so $\PD_k\{C(A\cup T;\al)\}=\PD_k\{C(A;\al)\}$ for $k\geq 1$.
\bs  
\end{problem}

Theorem~\ref{thm:tail} gave only sufficient conditions that guarantee the same 1D persistence under adding a tail.
Problem~\ref{pro:tail_necessary} asks to weaken these conditions.

\begin{problem}[necessary conditions for preserving persistence]
\label{pro:tail_necessary}
For each filtration from Definition~\ref{dfn:complexes}, find sufficient a necessary conditions on a cloud $A$ and its tail $T$ such that $\PD_k\{C(A\cup T;\al)\}=\PD_k\{C(A;\al)\}$ for $k\geq 1$.
\bs  
\end{problem}

\cite{oudot2020inverse} previously asked to find one point cloud for a given persistence: ``If a given persistence module does come from a point cloud, can that point cloud be computed effectively?''
Corollary~\ref{cor:PD=0} described a generic family of clouds $A\cup T$ that all have trivial persistence $\PD_1=\emptyset$.
The deeper problem below requires us to geometrically interpret persistence as an equivalence of clouds.

\begin{problem}[persistence as equivalence]
\label{pro:persistence}
Geometrically describe an equivalence relation on point clouds $A$. e.g. as transformations of the ambient space, whose classes are in a 1-1 correspondence with persistence diagrams $\PD_k\{C(A;\al)\}$ for $k\geq 1$.
\bs  
\end{problem}

Theorem~\ref{thm:tail} motivated comparisons of persistent homology with other isometry invariants of point clouds.
For finite sets of $m$ ordered points, a complete isometry invariant is a classical distance $m\times m$ matrix whose brute-force adaptation to unlabelled points requires $m!$ permutations.
The simpler collection of $\dfrac{m(m-1)}{2}$ pairwise distances (with repetitions) between $m$ unlabeled points is complete for sets in general position \cite{boutin2004reconstructing} but do not distinguish infinitely many non-isometric $m$-point sets for $m\geq 4$.
\medskip
 
The local distribution of distances \cite{memoli2011gromov} was recently studied under the name of the Pointwise Distance Distribution (PDD) for finite and periodic sets \cite{widdowson2022resolving}.
The completeness of the PDD is easy for finite sets in general position in $\bR^N$ \cite[Theorem 16]{widdowson2022average} and was recently extended to the much harder periodic case \cite[Theorem~4.4]{widdowson2022resolving}.
The PDD is conjectured to be complete for $N=2$ but cannot distinguish counter-examples \cite{pozdnyakov2020incompleteness} for $N=3$, which were classified by higher order invariants in appendix~C of the first version of \cite{widdowson2021pointwise} in 2021.
\medskip

The recent even stronger invariants \cite{kurlin2022computable, kurlin2024polynomial} were proved to be Lipschitz continuous  \cite{kurlin2023strength} and complete under rigid motion in any Euclidean space $\bR^N$ \cite{widdowson2023recognizing},  extended to metric spaces with measures \cite{kurlin2023simplexwise}.
The Lipschitz continuity is important for accurate predictions of material properties \cite{ropers2022fast,balasingham2024material,balasingham2024accelerating}.
\medskip
 
Another advantage of the PDD is its near-linear time based on a new algorithm for nearest neighbours \cite{elkin2023new}, which corrected gaps in the past proofs for cover trees \cite{elkin2022counterexamples}.
The actual speed is so fast that more than 200 billion pairwise comparisons of all 660K+ periodic crystals in the world's largest database of real materials were done within two days on a modest desktop.
This experiment detected physically impossible isometric duplicates whose underlying publications are investigated by five journals for data integrity \cite[section~7]{widdowson2022average}.
\medskip

More importantly, the above experiment justified the Crystal Isometry Principle  (CRISP) saying that all real periodic crystals have unique locations determined by their complete isometry invariants 
in a common Crystal Isometry Space continuously parametrised by complete isometry invariants.
Even if examples of periodic sets with the same PDD emerge, the slower isoset invariant is provably complete \cite{anosova2021isometry} and has continuous metrics \cite{anosova2022algorithms}.
\medskip

The second author proved the theoretical results and wrote the paper. 
The first author completed the experiments in section~\ref{sec:experiments} and reviewed the writing.
\backmatter

\bmhead{Acknowledgments}

We thank all reviewers, especially one anonymous reviewer for their helpful suggestions to  simplify Definition~\ref{dfn:long_wedge} and the proofs of Theorems~\ref{thm:long_wedges} and \ref{thm:tail}.
This work was supported by the £3.5M EPSRC grant `Application-driven Topological Data Analysis' (2018-2023),  the last author's Royal Academy of Engineering Fellowship `Data Science for Next Generation Engineering of Solid Crystalline Materials' (2021-2023, IF2122/186), the EPSRC New Horizons grant `Inverse design of periodic crystals' (2022-2024, EP/X018474/1) and the Royal Society APEX fellowship `New geometric methods for mapping the space of periodic crystals' (2023-2025, APX/R1/231152).
\medskip

\noindent
\textbf{Declaration}. 
The authors state that there is no conflict of interest.



\bibliographystyle{spmpsci}
\bibliography{APCT-D-22-00066final}


\end{document}